\documentclass[fleqn,12pt,twoside]{article}
\usepackage[headings]{espcrc1}

\readRCS $Id: espcrc1.tex,v 1.2 2004/02/24 11:22:11 spepping Exp $
\usepackage{graphicx}
\usepackage[figuresright]{rotating}

\newcommand{\AmS}{{\protect\the\textfont2
  A\kern-.1667em\lower.5ex\hbox{M}\kern-.125emS}}

\title{Scattering of Quark-Quasiparticles in the Quark-Gluon Plasma}

\author{M.~Mannarelli\address[MCSD]{Center for Theoretical Physics,
Laboratory for Nuclear Science and Department of Physics,
Massachusetts Institute of Technology, Cambridge, MA 02139}%
   \,\,and\,\, R.~Rapp\address{Cyclotron Institute and Physics Department,
Texas A\&M University, College Station, Texas 77843-3366} }

\begin{document}

\maketitle

\begin{abstract}
Employing a Brueckner-type many-body approach, based on a driving
potential extracted from lattice QCD, we study light quark
properties in a Quark-Gluon Plasma (QGP) at moderate temperatures,
$T\simeq$~1-2~$T_c$. The quark-antiquark $T$-matrix is calculated
self-consistently with pertinent quark self-energies. While the
repulsive octet channel induces quasiparticle masses of up to
150~MeV, the attractive color-singlet part exhibits resonance
structures which lead to quasiparticle widths of $\sim$200~MeV.
\end{abstract}

\section{Introduction}
Over the past few years the properties of the Quark-Gluon Plasma
(QGP) at moderate temperatures have received renewed interest. On
the one hand, experimental evidence from the Relativistic Heavy-Ion
Collider (RHIC) suggests the formation of a ``strongly interacting
QGP" to reconcile the phenomenological success of hydrodynamic
approaches with the inherent short thermalization times of
$\sim$0.5~fm/c. On the other hand, lattice QCD (lQCD)
computations~\cite{Asakawa:2002xj,Karsch:2003jg} indicate the
formation of mesonic bound (and/or resonance) states, which are also
supported by applications of lQCD-based
potentials~\cite{Kaczmarek:2003dp} in a Klein-Gordon
equation~\cite{Shuryak:2004tx}. To better understand the scattering
aspects of this problem, we have recently implemented lQCD-based
potentials into a selfconsistent scheme of quark-antiquark
$T$-matrix and quark selfenergies~\cite{Mannarelli:2005pz}, which we
will report on in this talk.

\section{Brueckner Approach}
\begin{figure}[th]
\begin{center}
\includegraphics[height=.3\textheight,width=.2\textheight,angle=-90]{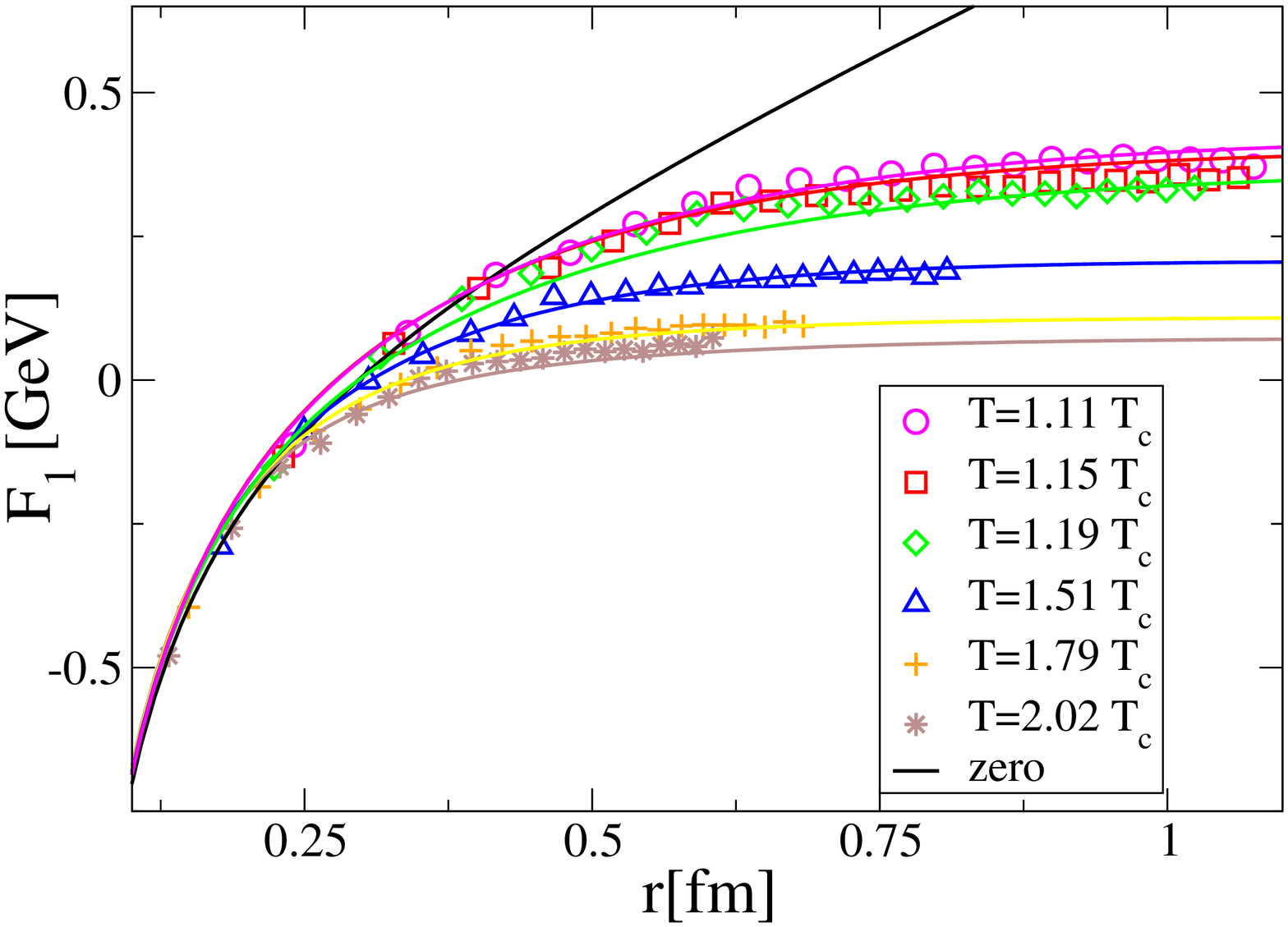}
\includegraphics[height=.3\textheight,width=.2\textheight,angle=-90]{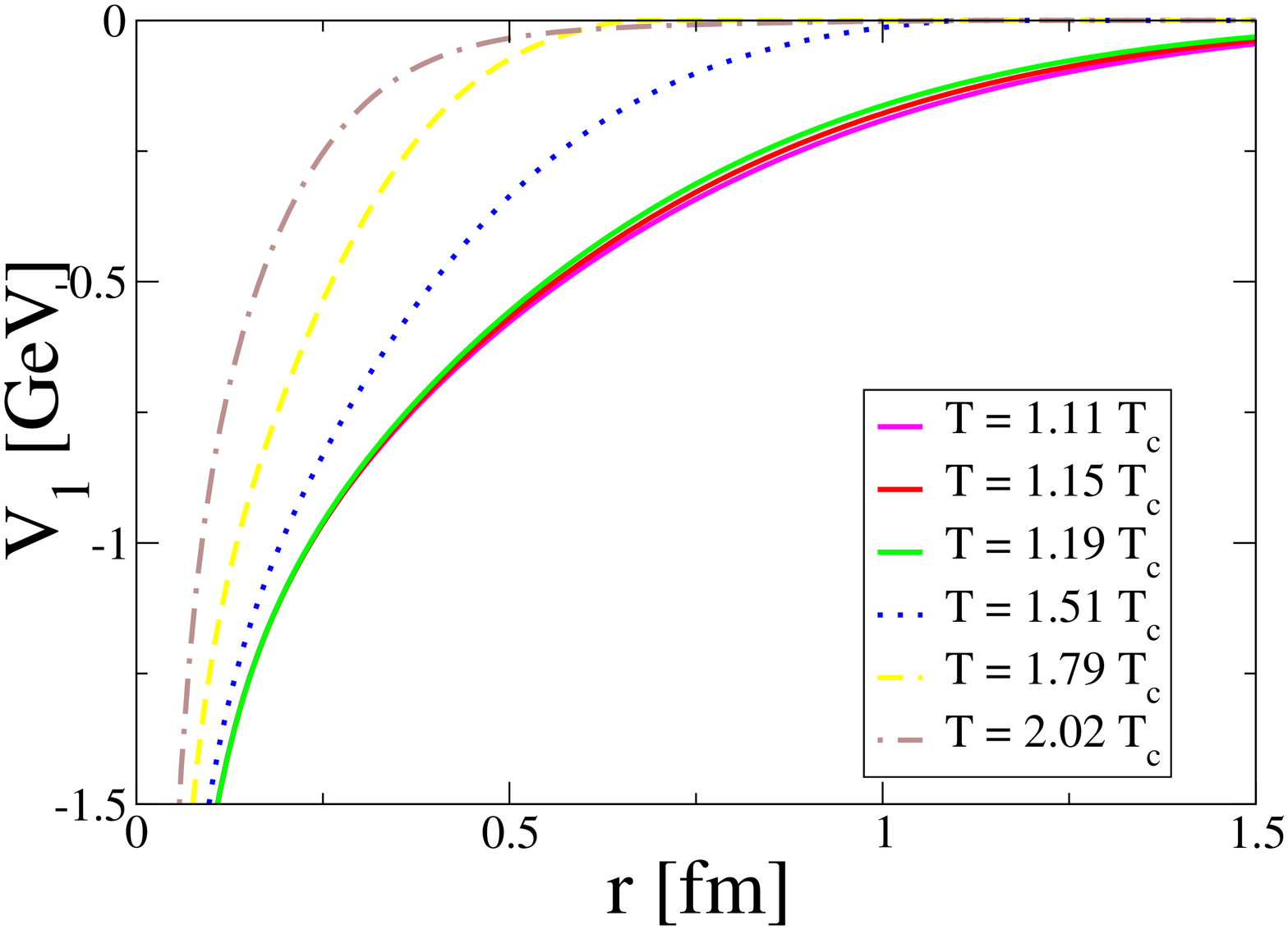}
\end{center}
\vspace{-0.8cm}
\caption{Left panel: color-singlet free energy from unquenched lQCD
simulations  for 6 different values of the
temperature (symbols) compared to our fit function,
Eq.~\ref{singlet}, represented by the various curves. Right panel:
corresponding potentials in the color-singlet channel obtained with
Eq.~\ref{V_1} for the same values of the temperature.}
\label{fig_lat}
\end{figure}

To obtain a driving term (potential) for a $q$-$\bar q$ scattering
equation we take recourse to lQCD calculations of the static free
energy for a (heavy) $Q$-$\bar Q$ pair. For temperatures
$T$$\simeq$~1.1-2~$T_c$, the unquenched singlet free
energy~\cite{Kaczmarek:2003dp} can be well reproduced by
\begin{equation}
F_1(r,T) = -\frac{\alpha}{r}e^{-a \mu(r,T) r} +
        \frac{\sigma}{\mu(r,T)}(1-e^{-\mu(r,T) r }) \, ,
\label{singlet}
\end{equation}
where $ \mu(r,T) = \frac{\sigma}{b}
{\rm exp(-0.3/r)} $ is a ``screening mass", $a$ and $b$ are
fitting functions (see \cite{Mannarelli:2005pz} for details), and
$\alpha=0.4$, $\sigma=1.2 \,{\rm GeV}^2$, cf.~left panel of
Fig.~\ref{fig_lat}.
The internal energy is obtained by
subtracting the entropy contribution to the free energy,
\begin{equation}
E_1\,=\,F_1-T\frac{d F_1}{d T} \,
\end{equation}
and the potential is the defined by subtracting the asymptotic value
of the internal energy (which is interpreted as a mass term),
\begin{equation} V_1(r,T)=E_1(r,T) -E_1(\infty,T) \, . \label{V_1}
\end{equation}
cf.~right panel of Fig.~\ref{fig_lat}. We also consider the (repulsive)
color-octet channel assuming that the potential follows the leading-order
result of perturbation theory,
$F_8 = - \frac{1}{8} F_1 $. Relativistic corrections are
included via a velocity-velocity interaction~\cite{Brown:1952ph}.
\begin{figure}[!b]
\vspace{-0.2cm}
\begin{center}
\includegraphics[height=.2\textheight,width=.4\textheight,angle=0]
{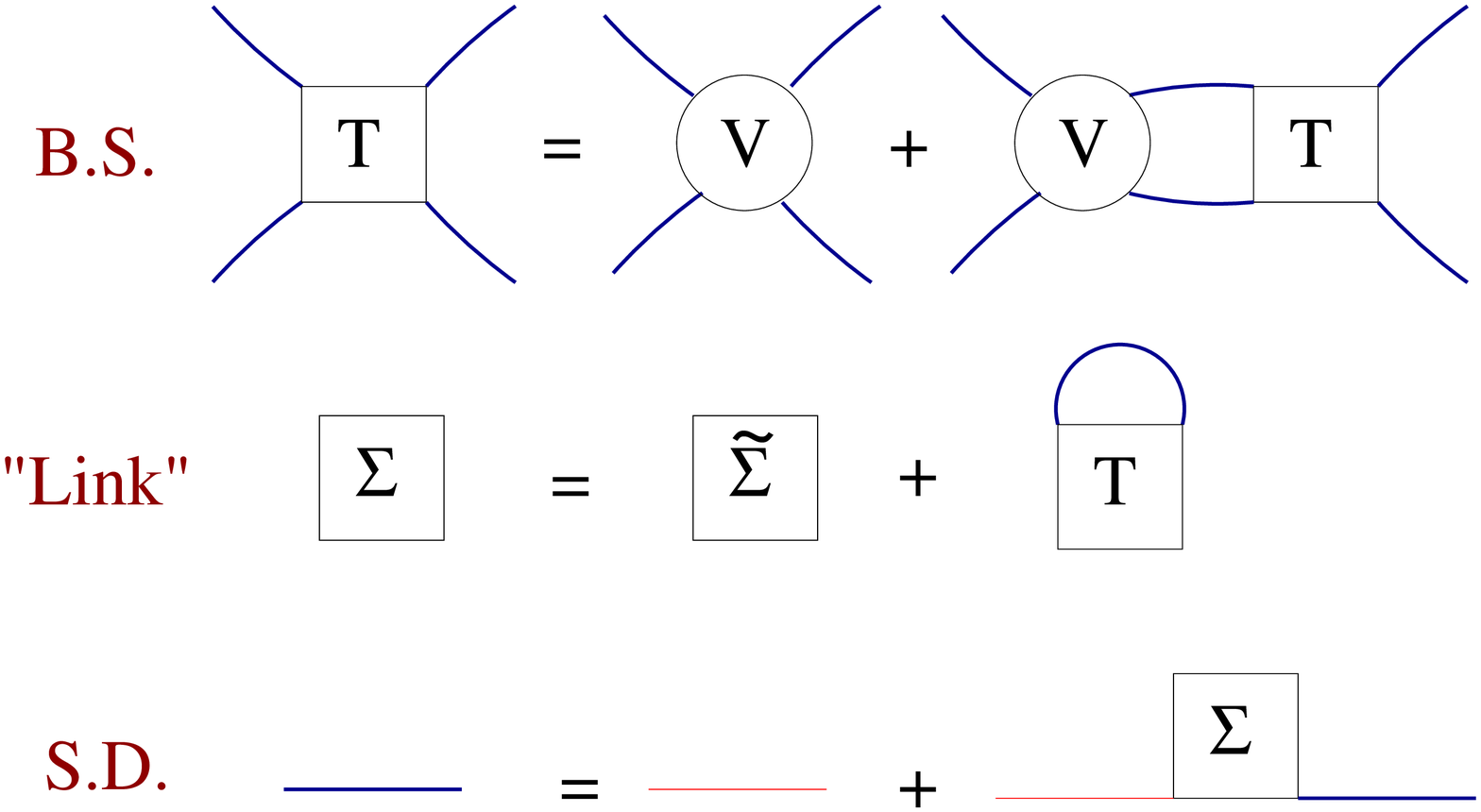}
\end{center}
\vspace{-0.8cm} \caption{Schematic representation of the
self-consistency problem, Eqs.~(\ref{self}). Upper panel (B.S.):
Bethe-Salpeter equation, middle panel (``Link"): single-quark
selfenergy, lower panel (S.D.): Schwinger-Dyson equation for the
quark propagator; thick (blue) lines: full quark propagators, thin
(red) lines: bare quark propagators.} \label{fig_BSSD}
\end{figure}

The quark-antiquark interactions in the QGP are evaluated in the
$T$-matrix approach, as is well known from the nuclear many-body
problem. In relativistic field theory, the starting point is a
system of coupled equations,
\begin{equation}
 T = K + \int K S S T \, ,  \quad S =  S_0 + S_0 \Sigma S \, , \quad
\Sigma = \tilde \Sigma +\int\! T S \label{Sigma0} \, ;
\label{self}
\end{equation}
the first is a 4-dimensional Bethe-Salpeter equation
($K$: interaction kernel), the second a Schwinger-Dyson equation for
the single-quark propagator, $S$ ($S_0$: vacuum propagator),
where the medium effects
are encoded in a self-energy  $\Sigma$, which, via the third equation,
depends on the two-body $T$-matrix ($\tilde\Sigma$ represents
a contribution due to quark interactions with thermal
gluons which we treat as a ``gluon-induced" mass
$m$ in the quark dispersion law~\cite{Mannarelli:2005pz}).
Eqs.~(\ref{self}) constitute a self-consistency problem which
is diagrammatically illustrated in Fig.~\ref{fig_BSSD}. After employing
an appropriate non-relativistic reduction scheme in line with the
potential approximation, we have solved Eqs.~(\ref{self}) by numerical
iteration for a vanishing baryon chemical potential implying
identical results for quarks and antiquarks.

\section{Selfconsitent Scattering Amplitudes and Selfenergies}
\begin{figure}
\begin{center}
\noindent
\includegraphics[height=2.in,angle=-90]{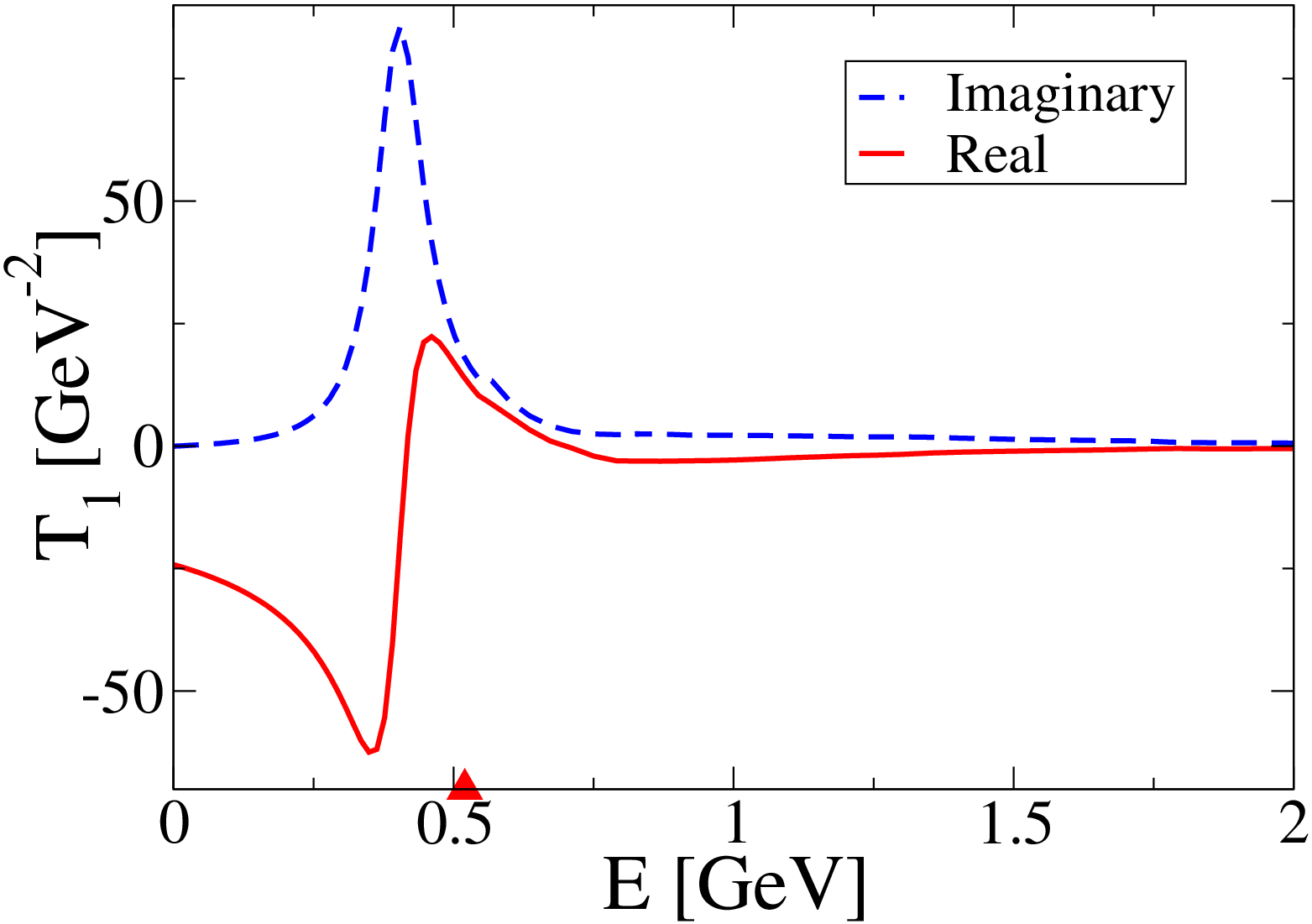}
\includegraphics[height=2.in,angle=-90]{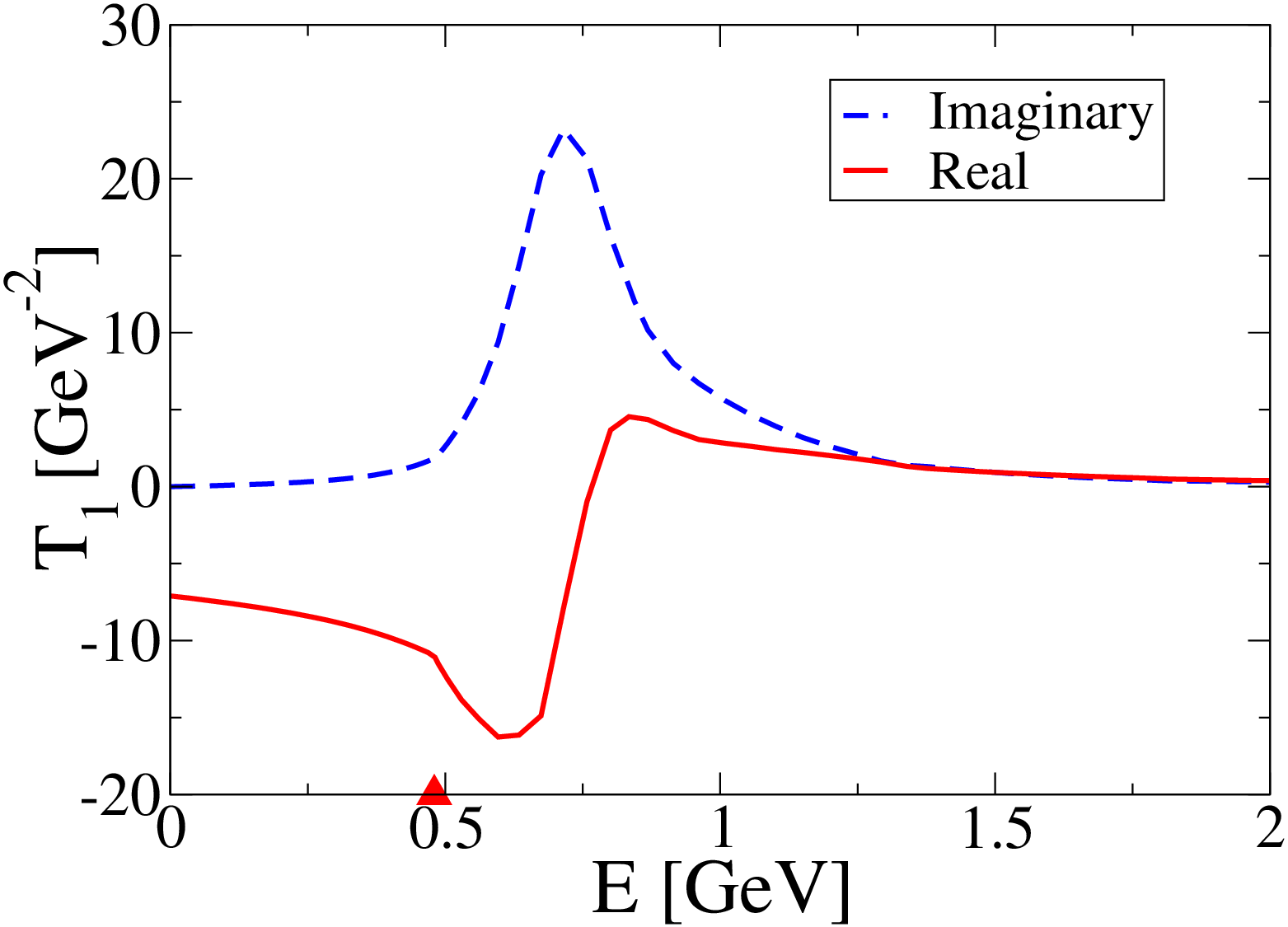}
\includegraphics[height=2.in,angle=-90]{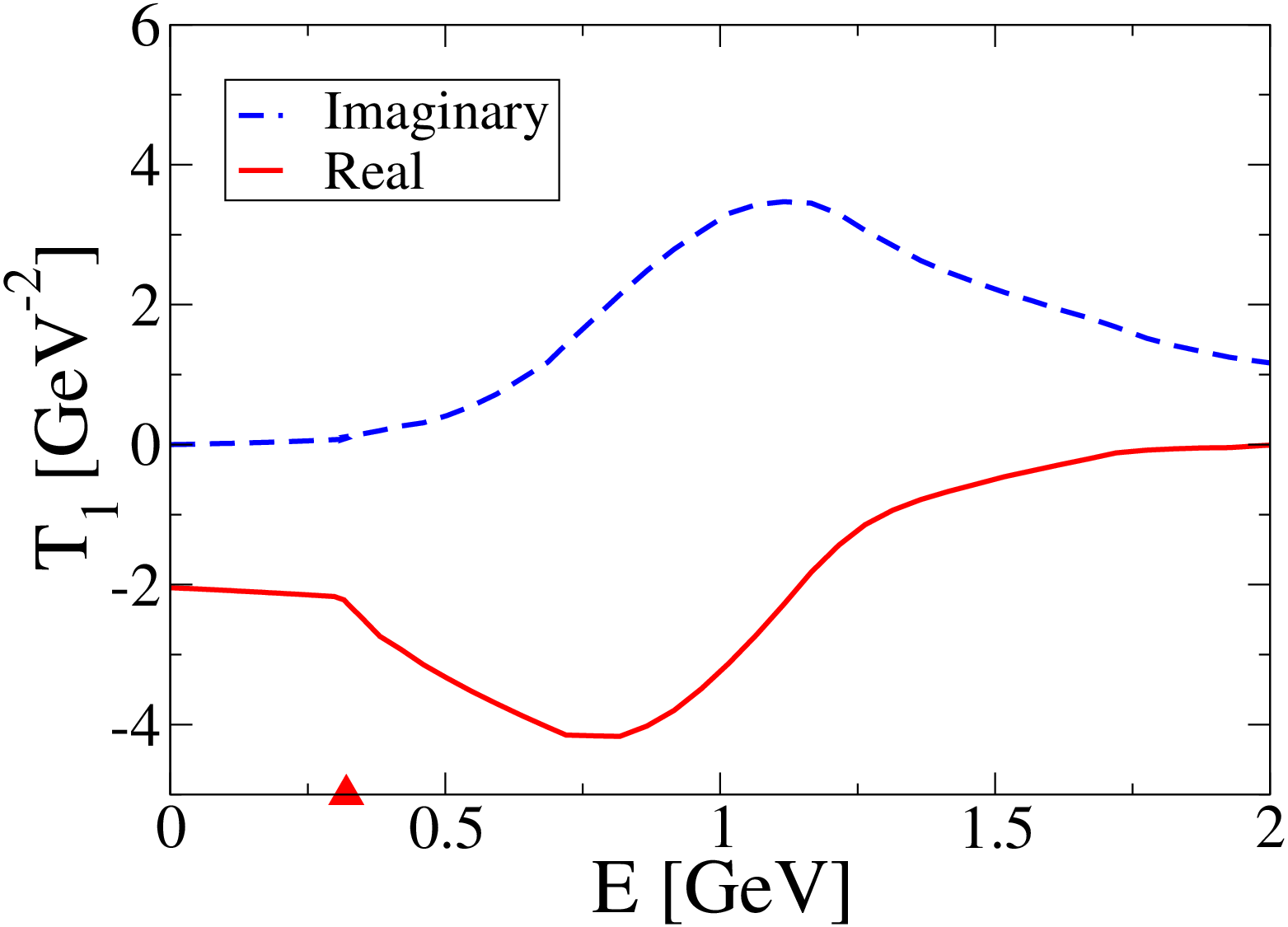}
\includegraphics[height=2.0in,angle=-90]{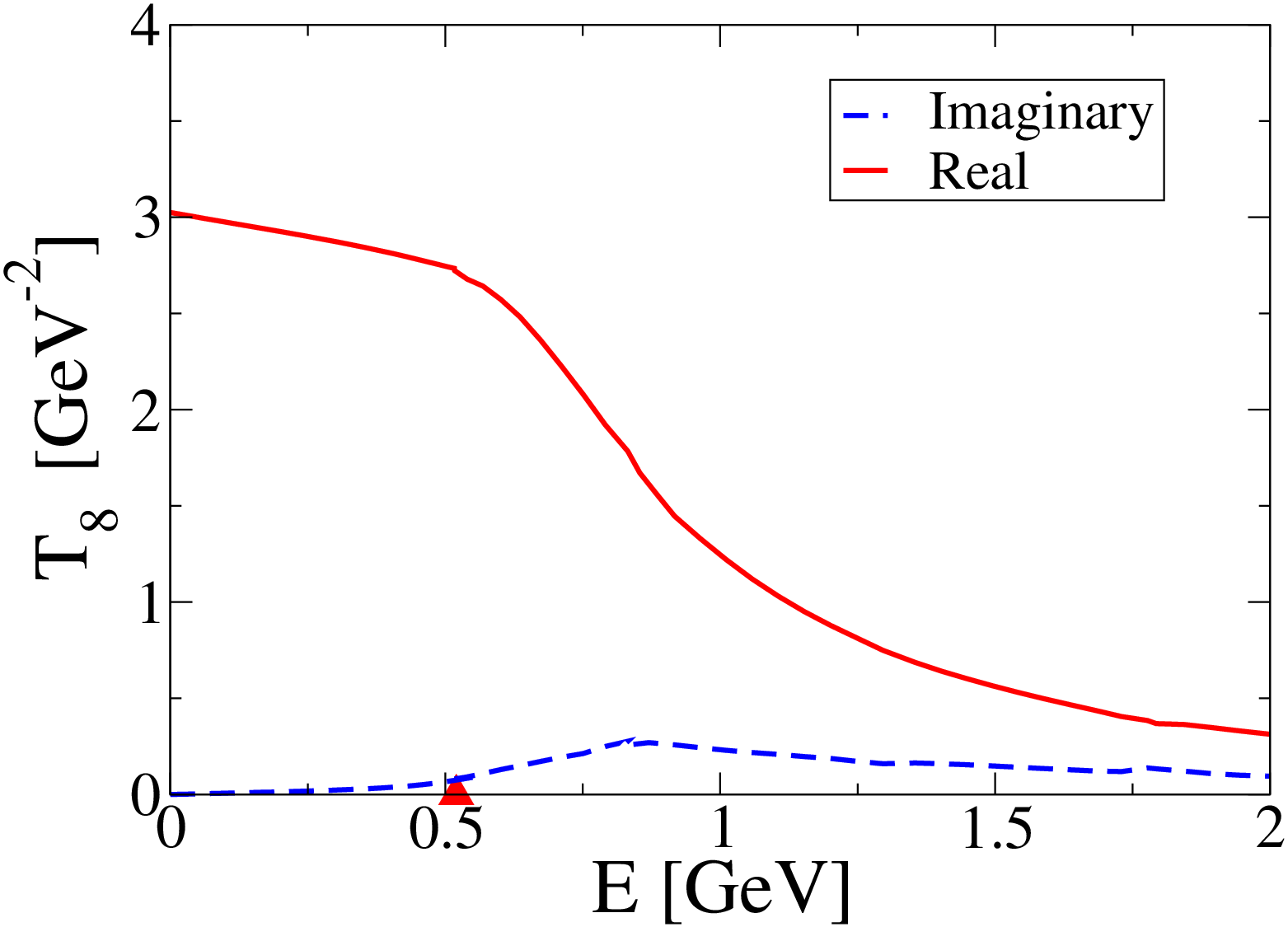}
\includegraphics[height=2.0in,angle=-90]{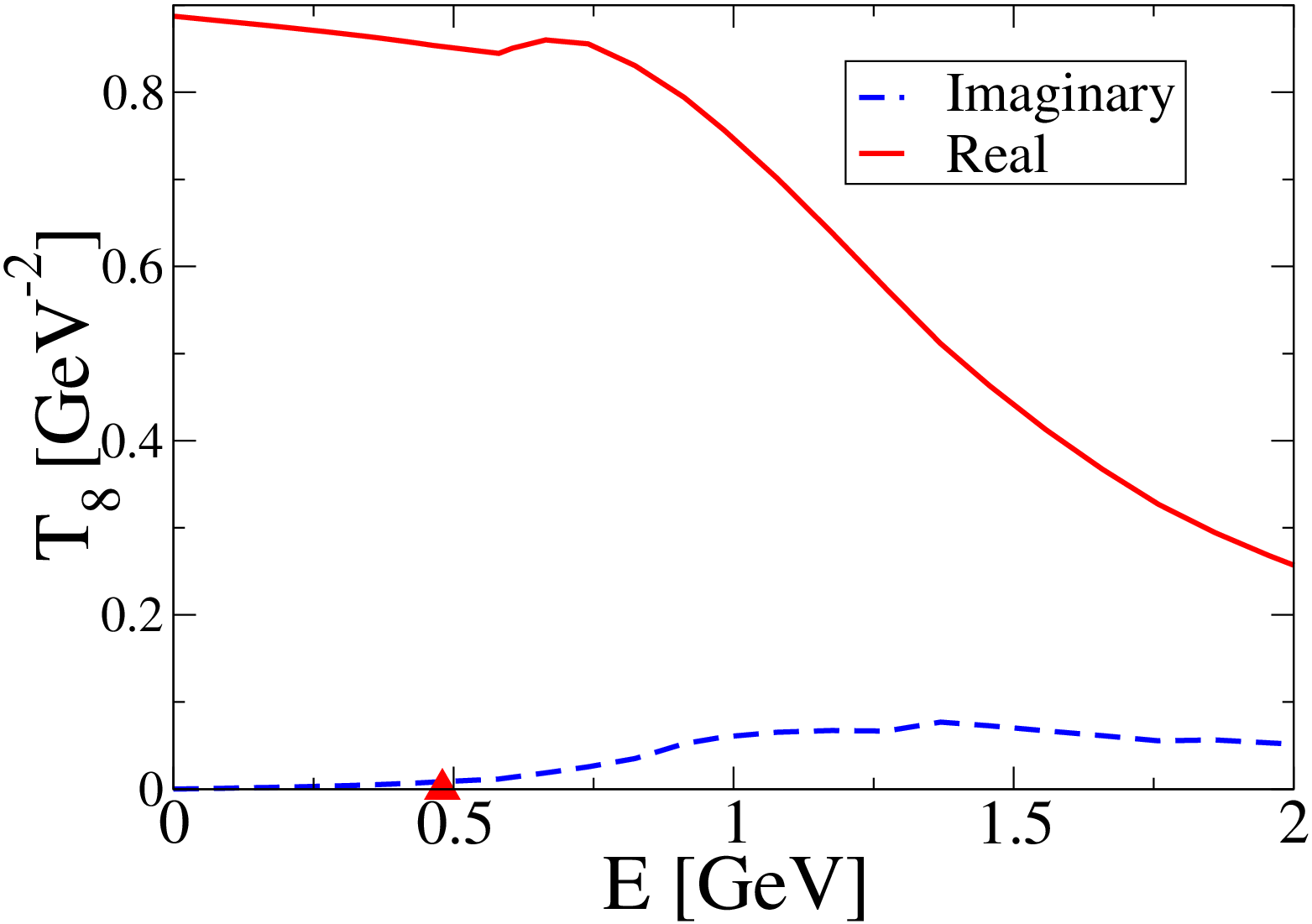}
\includegraphics[height=2.0in,angle=-90]{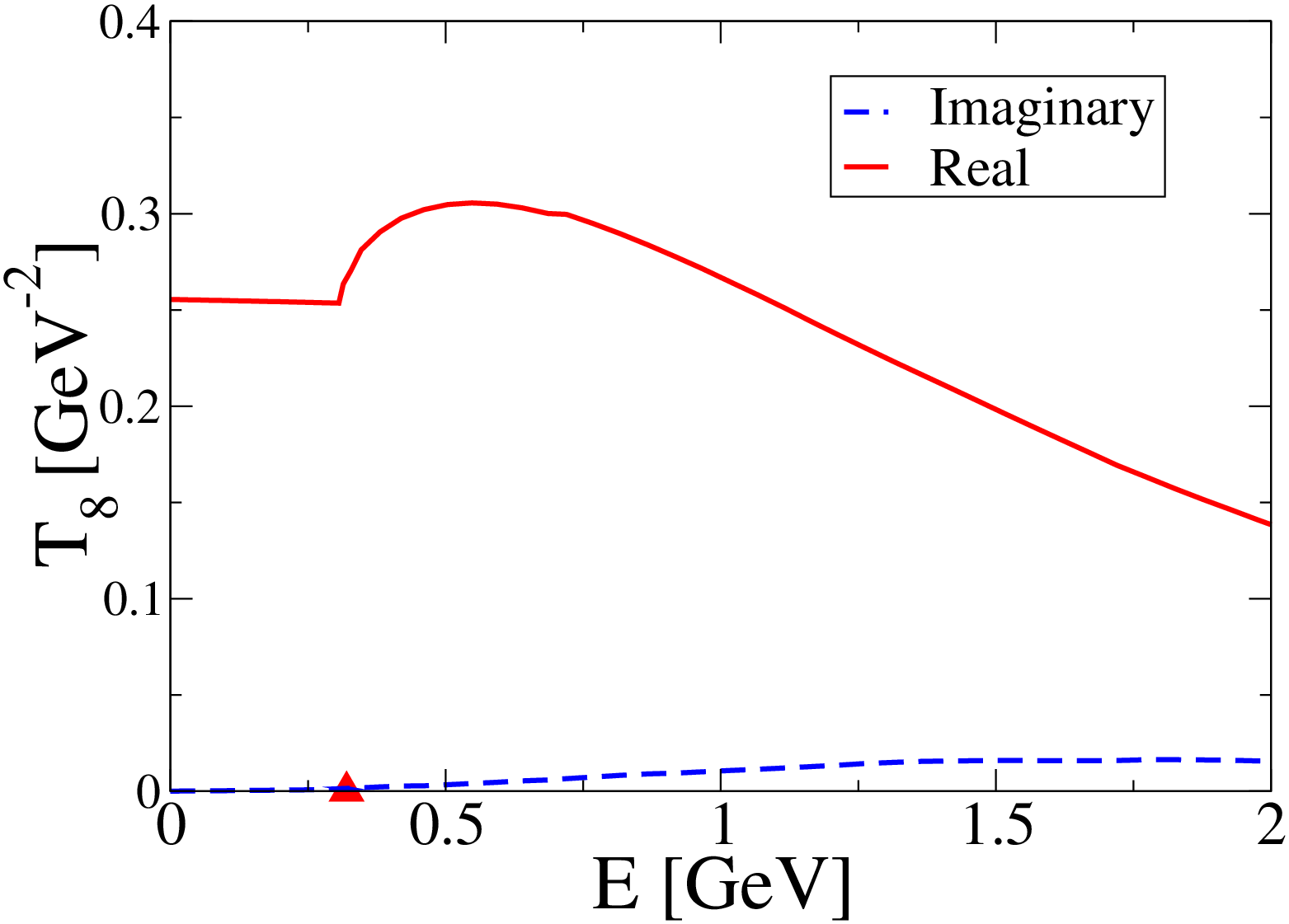}
\end{center}
\vspace{-0.8cm}
\caption{Real (full red line) and imaginary part (dashed blue line)
of the $T$-matrix in the color-singlet (upper panel) and
color-octet (lower panels) channels at temperatures
$T$=1.2, 1.5 and 1.75~$T_c$ (left, middle and right
panel, respectively) as a function of $q\bar q$ $CM$ energy $E$,
with $m$=0.1~GeV. Red triangles on the $x$-axis indicate the
threshold energy.}
\label{figTm1}
\end{figure}

Fig.~\ref{figTm1} summarizes our results for the selfconsistent
on-shell $T$-matrices (real and imaginary parts)
with a ``gluon-induced" mass term of $m$=0.1~GeV for 3 different
temperatures.
At $T$=1.2~$T_c$, the color-singlet $T$-matrix (upper left panel of
Fig.~\ref{figTm1}) exhibits a relatively narrow bound state located
significantly below the $q$-$\bar q$ threshold energy,
$E_{thr}\simeq0.52$~GeV. With increasing temperature (middle and right
panels), this state moves above threshold (i.e., becomes a resonance)
and broadens substantially, being essentially melted at $T$=1.75~$T_c$.
These results are in qualitative agreement with computations of
mesonic spectral functions in (quenched)
lQCD~\cite{Asakawa:2002xj,Karsch:2003jg}. The color-octet $T$-matrix
(lower panels of Fig.~\ref{figTm1}) depends rather
smoothly on $CM$ energy, decreasing in strength with temperature with
the (repulsive) real part being much larger in magnitude
than the imaginary part.

\begin{figure}
\begin{center}
\includegraphics[height=.2\textheight,angle=-90]{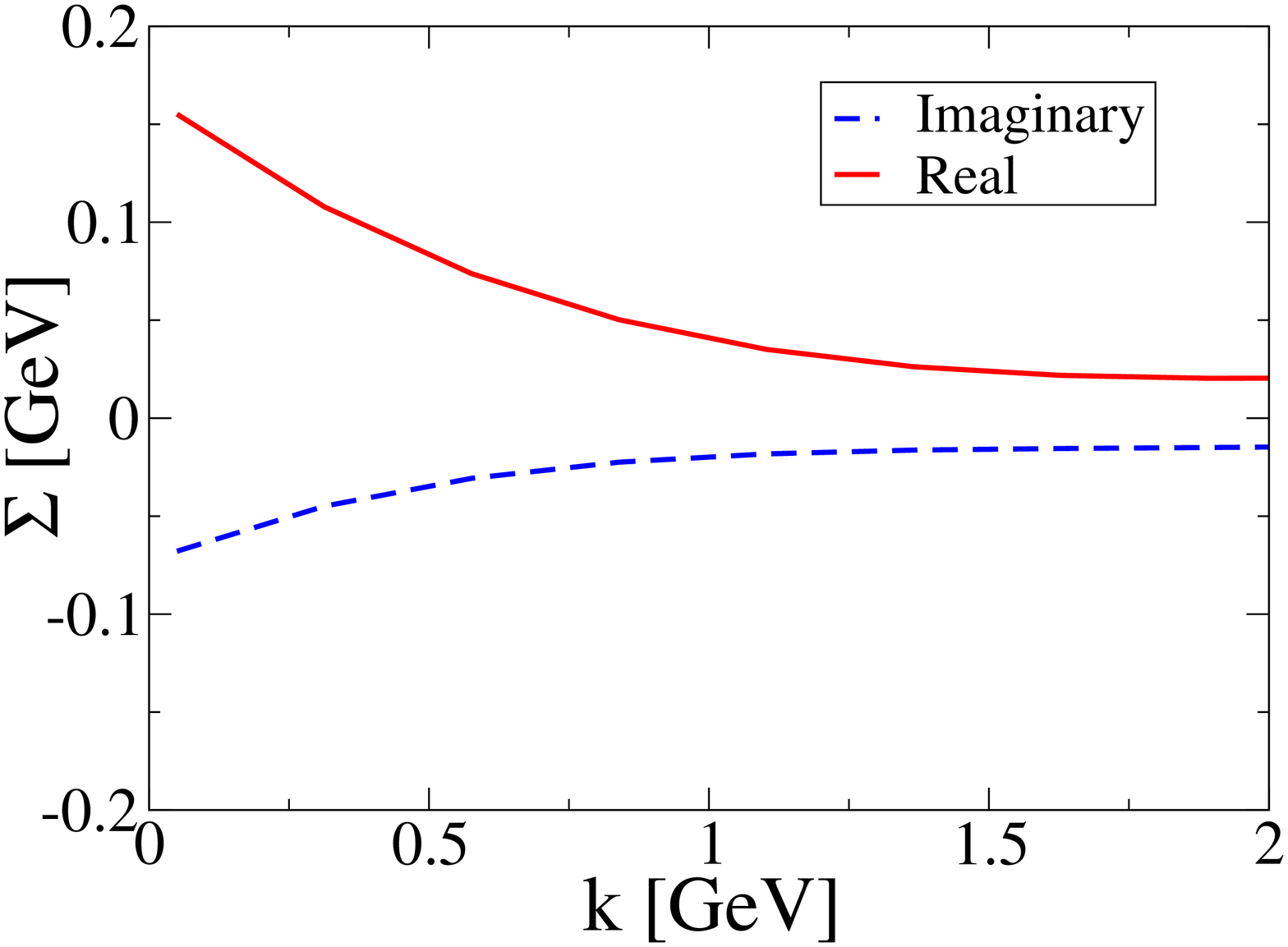}
\includegraphics[height=.2\textheight,angle=-90]{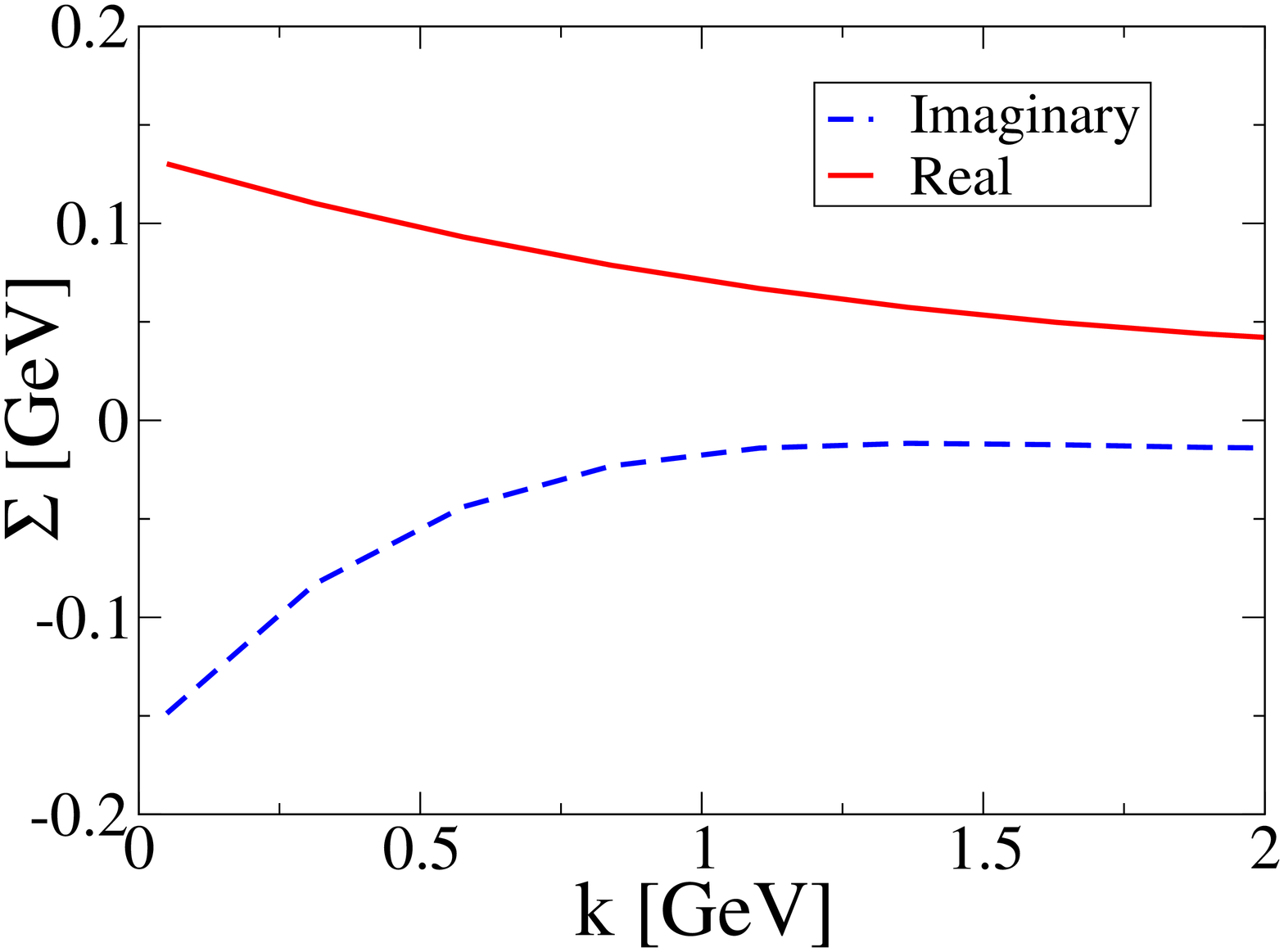}
\includegraphics[height=.2\textheight,angle=-90]{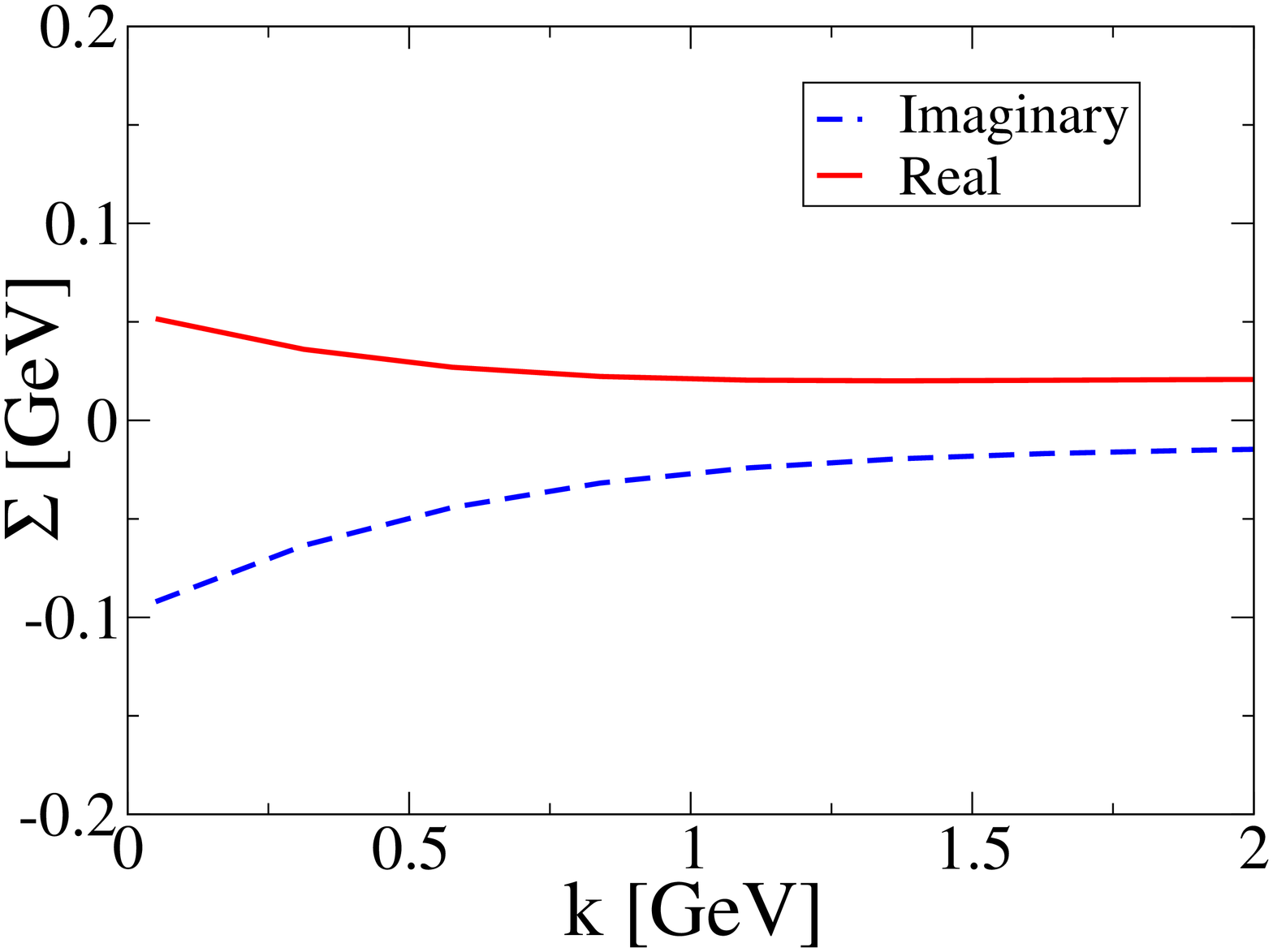}
\end{center}
\vspace{-0.8cm}
\caption{Real (solid line, red) and imaginary (dashed line, blue)
part of the on-shell quark self-energy as a function of 3-momentum
at temperatures $T$=1.2, 1.5 and 1.75~$T_c$ (left, middle and right
panel, respectively), with $m$=0.1~GeV.}
\label{figSelf1}
\end{figure}
Fig.~\ref{figSelf1} displays the self-consistent on-shell quark
self-energies for the same temperatures as in Fig.~\ref{figTm1}.
Both real and imaginary parts are
smooth functions of the quark 3-momentum with maximal values at
$k$=0. The positive real part mostly arises due to the repulsive
color-octet $T$-matrix with a degeneracy factor of
8 times the singlet one. Pertinent (nonperturbative) thermal quark
masses amount to 150~MeV at small momenta for $T$=1.2-1.5~$T_c$,
decreasing to $\sim$50~MeV at 1.75~$T_c$.
With the underlying ``gluon-induced" mass term of $m$=100~MeV, the
total thermal mass, $m + \Sigma_R$, adds to 150-250~MeV.
The imaginary part is chiefly generated by resonant scattering
in the color-singlet channel, translating into quasiparticle
widths of $\sim$200~MeV at low momenta.

\section{Summary and Conclusions}
Anti-/quark self-energies and scattering amplitudes have been
evaluated within a self-consistent Brueckner approach using
temperature-dependent lQCD-based potentials. Our calculations
support the notion of mesonic resonances in the QGP at temperatures
below 2~$T_c$. Pertinent quasiparticles masses and widths are
appreciable ($\sim$100-200~MeV) and comparable, qualitatively
suggesting that the QGP could be in a liquid-like
regime~\cite{Thoma:2004sp}.

\vspace{0.5cm}

\noindent
{\bf Acknowledgement} \\
One of us (RR) is supported in part by a U.S. National Science Foundation
CAREER award under grant PHY-0449489. One of us (MM) is supported in part
by the U.S. Department of Energy under cooperative research agreement
\#DE-FC02-94ER40818.




\end{document}